\documentstyle[epsfig]{acs}
\def\S{{\zeta}}

\begin{document}

\title[Forecasting price increments with NN]{
       Forecasting price increments \\ using an artificial Neural Network
}

\author[Filippo Castiglione]{FILIPPO CASTIGLIONE \\
Center for Advanced Computer Science, University of Cologne\\
ZPR/ZAIK, Weyertal 80, D - 50931 K\"oln, Germany\\
{\tt filippo@zpr.uni-koeln.de}
}
\pagerange{\pageref{firstpage}--\pageref{lastpage}}
\label{firstpage}
\maketitle

\begin{abstract}
{\em ABSTRACT\/}. 
Financial forecasting is a difficult task due to the intrinsic complexity
of the financial system. 
A simplified approach in forecasting is given by ``black box''
methods like neural networks that assume little about the structure of the
economy.
In the present paper we relate our experience using neural nets as financial
time series forecast method.
In particular we show that a neural net able to forecast the sign of
the price increments with a success rate slightly above 50 percent
\emph{can} be found.
Target series are the daily closing price of different assets and indexes
during the period from about January 1990 to February 2000.

\end{abstract}

\section{Introduction}
\noindent
Forecasting future values of an asset gives, besides the straightforward 
profit opportunities, indications to compute various interesting 
quantities such as the price of derivatives (complex financial products) or the 
probability for an adverse mode which is the essential information when assessing and 
managing the risk associated with a portfolio investment.

Forecasting the price of a certain asset (stock, index, foreign currency, etc.)
on the ground of available historical data, corresponds to the well known problem 
in science and engineering of time series prediction.
While many time series may be approximated with a high degree of confidence, financial
time series are found among the most difficult to be analyzed and predicted.
This is not surprising since the dynamics of the markets following at least the 
semi-strong EMH should destroy any easy method to estimate future activities 
using past informations.

Among the methods developed in \emph{Econometrics} as well as other 
disciplines \footnote{
see the vast bibliography with more than 800 entries at\\
{\tt www.stern.nyu.edu/~aweigend/Time-Series/Biblio/SFIbib.html} reported 
from \cite{vastbib}}, the artificial \emph{Neural Networks} (NN) are being used 
by ``non-orthodox'' 
scientists as non-parametric regression methods \cite{Campbell:97,LNCS:1524:98}.
They constitute an alternative to nonparametric regression methods like
\emph{kernel regression} \cite{Campbell:97}.
The advantage of using a neural network as non linear function approximator is that it 
appears to be well suited in areas where the mathematical knowledge of the stochastic 
process underlying the analyzed time series is unknown and quite difficult to be 
rationalized.
Besides, it is important to note that the lack of linear 
correlations in the financial price series and the already accepted evidence of an
underlying process different from i.i.d. noise point out to the existence of 
higher-order correlations or non-linearities. 
It is this non-linear correlation that the neural net may eventually catch during its
learning phase. If some macroscopic regularities, arising from the apparently chaotic 
behaviour of the large amount of components are present, then a well trained net
could identify and ``store'' them in its distributed knowledge representation system 
made by units and synaptic weights \cite{LNCS:1524:98,Refenes:97}.

In the following we will see that a well suited NN for each of a set of
price time series showing a ``surprising'' rate of success in predicting the 
\emph{sign} of the price change on a daily base \emph{can} be found.
Not less interesting, we will see that the foretold regularities in the time series
seem to be more present on larger time scale than on high frequency data,
as the performance of the net degrades if we go from monthly to minutes data.

\begin{figure}[htbp] 
 \begin{center}
  \leavevmode
  \epsfig{file=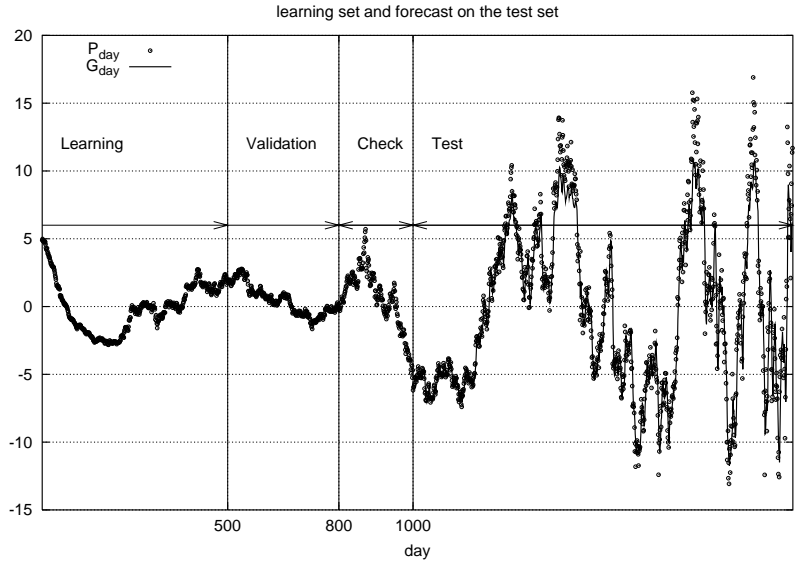,width = 0.9\textwidth}
  \caption[Learning, Validation, Check and Test data sets]
     { 
     Each time series is divided in four data sets:
     learning, validation, checking and testing (see text for explanation).
     A difficulty arise from the fact that the oscillations in the test set
     are much more pronounced than in the learning set.
     In figure, daily closing price of Intel Corp.}
     \label{fsets}
 \end{center}
\end{figure} 

\section{Multi-layer Perceptron}

\emph{Multi-layer perceptrons} (MLP) are the neural nets usually referred to
as function approximators.
A MLP is a generalization of Rosenblatt's
\emph{perceptron} (1958); $n_i$ input units, $n_h$ hidden and $n_o$ output units
with all feed forward connections between adjacent layers (no intra-layer
connections or loops). Such net's topology is specified as $n_i$-$n_h$-$n_o$.

A NN may perform various tasks connected to classification problems. Here we
are mainly interested in exploiting what is called the \emph{universal
approximation property}, that is, the ability to approximate any nonlinear function
to any arbitrary degree of accuracy with a suitable number of hidden units
\cite{White:92,Cybenko:89}.

The approximation is performed finding the set of weights connecting the units.
This can be done with one of the available methods of non-parametric estimation
techniques like \emph{nonlinear least-squares}.
In particular we choose \emph{error back propagation} which is probably the
most used algorithm to train MLPs \cite{Rumelhart:86a,Rumelhart:86b}. 
It is basically a gradient descent algorithm of the error computed on a 
suitable learning set. A variation of it use \emph{bias}, \emph{terms} and 
\emph{momentum} as characteristic parameters. 
Moreover we fixed the learning rate $\eta=0.05$, the momentum $\beta=0.5$ and the 
usual sigmoidal as nonlinear activation function.

\section{Detrending analysis}

We have trained the neural nets on ``detrended'' time series.
The detrending analysis was performed to mitigate the unbalance 
between the \emph{learning set}, and the \emph{test set}. 
In fact, subdividing the available data in learning set and testing set as 
specified in the following section (have a look at figure \ref{fsets}), 
we train the nets on a data set corresponding to a periods much back in time 
while we test the nets on data set corresponding to the most recent period of time. 
This problem is know in literature as \emph{noise/nonstationarity tradeoff} 
\cite{LNCS:1524:98}.

It is known that during the last ten years the American market has noticeably changed 
in that almost all the titles connected to the information technology have not only 
jumped to record values but also the fluctuations of price 
today are much stronger than before 
\footnote{
$P_t$ is what
we use to train our nets. Considering $\log(P_t)$ instead of $P_t$ would
mitigate the problem but it would introduce further nonlinearities
}.
Ignoring this fact would lead to a mistake because the net would not learn 
the characteristics of the ``actual situation''.

\begin{figure} 
 \begin{center}
 \leavevmode
 \epsfig{file=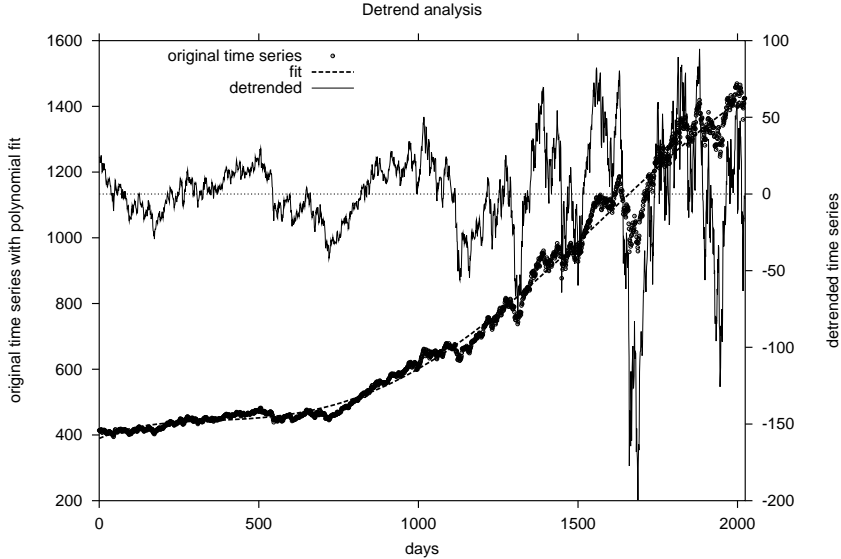,width = 0.9\textwidth}
 \caption[WHAT]{
        S\&P500 detrended time series. The plot shows the original series, the
        polynomial fit and the resulting detrended time series obtained
        just by difference between the original and the fitting curve.
        The detrended time series consist of 2024 points.}
        \label{fdetrend}
 \end{center}
\end{figure}

To detrend a time series we performed a {nonlinear least squares fit} using 
the \emph{Marquardt-Levenberg} algorithm \cite{Campbell:97,Press:94} 
with a polynomial of sixth degree. Then we just computed the difference of the series
with the fitting curve.
For each time series considered we ended up with a detrended series composed by 2024 
points corresponding to the period from about January 1990 to February 2000. 
For example, the plot in figure \ref{fdetrend} shows the detrended time series of
the index S\&P500 along with the original series and the polynomial fit.

We choose daily closing for 3 indexes and 14 assets historical series
on the NYSE and Nasdaq. In particular the assets were chosen among the most 
active companies in the field of information technology.

\section{Determining the net topology}
\label{dtnt}

One of the primary goals in training neural networks is to ensure that the
network will perform well on data that it has not been trained on (called
"generalization"). The standard method of ensuring good generalization is to
divide our training data into \emph{multiple data sets}. The most common data sets
are the \emph{learning} $L$, \emph{cross validation} $V$, and \emph{testing} $T$ 
data sets. While the learning data set is the data that is actually used to 
train the network the usage of the other two may need some explanation.

Like the learning data set, the cross validation data set is also used by
the network during training. Periodically, while training on the learning
data set, the network is tested for performance on the cross validation set.
During this testing, the weights are not trained, but the performance of the
network on the cross validation set is saved and compared to past values. If
the network is starting to overtrain on the training data, the cross
validation performance will begin to degrade. Thus, the cross validation
data set is used to determine when the network has been trained as well as
possible without overtraining (e.g., maximum generalization).

Although the network is not trained with the cross validation set, it uses
the cross validation set to choose a "best" set of weights. Therefore, it is
not truly an \emph{out-of-sample} test of the network. For a true test of the
performance of the network the testing data set $T$ is used. This data set is
used to provide a true indication of how the network will perform on new data.

\begin{figure} 
 \begin{center}
 \begin{picture}(0,0)%
   \epsfig{file=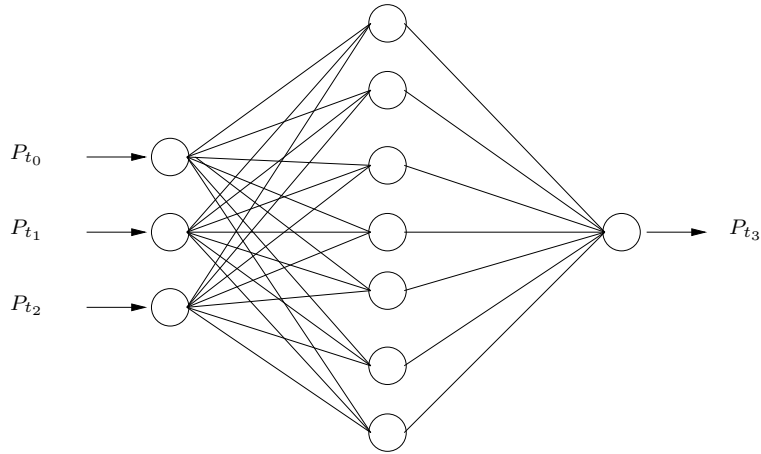}%
   \end{picture}%
   \setlength{\unitlength}{2763sp}%
   \begingroup\makeatletter\ifx\SetFigFont\undefined%
   \gdef\SetFigFont#1#2#3#4#5{%
     \reset@font\fontsize{#1}{#2pt}%
     \fontfamily{#3}\fontseries{#4}\fontshape{#5}%
     \selectfont}%
   \fi\endgroup%
   \begin{picture}(6450,4025)(751,-4211)
   \put(751,-1561){\makebox(0,0)[lb]{\smash{\SetFigFont{8}{9.6}{\rmdefault}{\mddefault}{\updefault}$P_{t_0}$}}}
   \put(751,-2236){\makebox(0,0)[lb]{\smash{\SetFigFont{8}{9.6}{\rmdefault}{\mddefault}{\updefault}$P_{t_1}$}}}
   \put(751,-2911){\makebox(0,0)[lb]{\smash{\SetFigFont{8}{9.6}{\rmdefault}{\mddefault}{\updefault}$P_{t_2}$}}}
   \put(7201,-2236){\makebox(0,0)[lb]{\smash{\SetFigFont{8}{9.6}{\rmdefault}{\mddefault}{\updefault}$P_{t_3}$}}}
   \end{picture}
 \caption[A Multi-Layer Perceptron]{
    A three layer perceptron $3-7-1$ with three inputs, seven hidden and
    one output units.}
    \label{nn}
 \end{center}
\end{figure}

In figure \ref{nn}, an example of MLP with $n_i=3$, $n_h=7$ and one output 
unit takes $P_{t_0}, P_{t_1}, P_{t_2}$ in input and gives the successive value 
$P_{t_3}$ as forecast. The number of free parameters is given by the number of 
connections between units $(n_i+n_o)\cdot n_h$.

While the choice of one output unit comes from the straightforward definition of 
the problem, a crucial question is ``how many input and hidden units should we choose?''.
In general there is no way to determine apriori a good network topology.
It depends critically on the number of training examples and the complexity of the
time series we are trying to learn.
To face this problem a large number of methods are being developed (recurrent networks,
model selection and pruning, sensitivity analysis \cite{LNCS:1524:98}), 
some of which follow the evolution's paradigm (Evolutionary Strategies and 
Genetic Algorithm).

Because we have observed a critical dependence of the performance of the net
from $n_i$ and $n_h$, and to avoid the great complexity of more powerful strategies 
\cite{LNCS:1524:98}, we ended up with the decision to explore all the possible  
combinations of $n_i$-$n_h$ in a certain range of values. 
Our ``brute force'' procedure consists of training nets of different topologies 
(varying $2\le n_i \le 15$ and $2\le n_h \le 25$) and observe their performance. 
More precisely we select good nets on the basis of the mean square error 
(see eq(\ref{eqe})) computed on 200 points out of the sample set constituting the test 
set. Thus, besides the separation in Learning-Validation-Testing of our time series, 
we further distinguish a subset from the Testing set: the \emph{Checking} $C$ 
(see fig.~\ref{fsets}).
The reason is that while we train the net to interpolate the time series
(minimizing the mean square error) we finally extrapolate to forecast the 
\emph{sign of the increments} (to be defined later).

To assess the efficiency of the learning and to discard bad trained nets during the 
search procedure we use the \emph{mean square error} $\epsilon$ defined as
\begin{equation}
  \epsilon = \frac{1}{\sigma} \cdot \frac{1}{|C|} \sum_{t\in C} \left(G_t-P_t\right)^2
\label{eqe}
\end{equation}
where $P_t$ is the price value, $G_t$ is the forecasted value at time $t\in C$ and
$\sigma$ is the standard deviation of the time series.
For good forecasts we will have small positive values of 
$\epsilon$ ($1\gg\epsilon\ge 0$).

We set the threshold 0.015 to discriminate good from bad nets. Only those nets for which
$\epsilon \le 0.015$ are further tested for sign prediction.

In summary, first we learn on set $L$, and through validation $V$ we find when to stop 
learning; then through check on $C$ we see if the learning process worked well, and in 
case it did, we make predictions in the test phase on set $T$ for "future" 
(i.e. previously unused) price changes and compare them with reality.

\section{Stopping criteria} 
To avoid overfitting and/or very slow convergence of the training phase, 
the stopping criteria is determined by the following three conditions, one of which
is sufficient to end the training phase (early stopping):

\begin{enumerate}
\item Stopping is assured within 5000 iterations of cross validation 
(see section \ref{dtnt});
\item during cross validation the mean square error on the validation set $V$ is 
computed as $\varepsilon_V = \frac{1}{2} \sum_{t\in V} \left(G_t-P_t\right)^2$; 
during training $\varepsilon_V$ should decrease, so a 
stopping condition is given if $\varepsilon_V$ increase again more than 20\% of 
the minimum value reached up to then;
\item learning is also stopped if $\varepsilon_V$ reaches a plateau; this is 
tested during cross validation averaging 1000 successive values of $\varepsilon_V$ 
and checking if the actual value is above this average.

\end{enumerate}

\def\piccolo{\fontsize{0.2cm}{0.24cm}\selectfont}
\section{Results}

\begin{figure}[htbp] 
  \begin{center}
  \leavevmode
  \epsfig{file=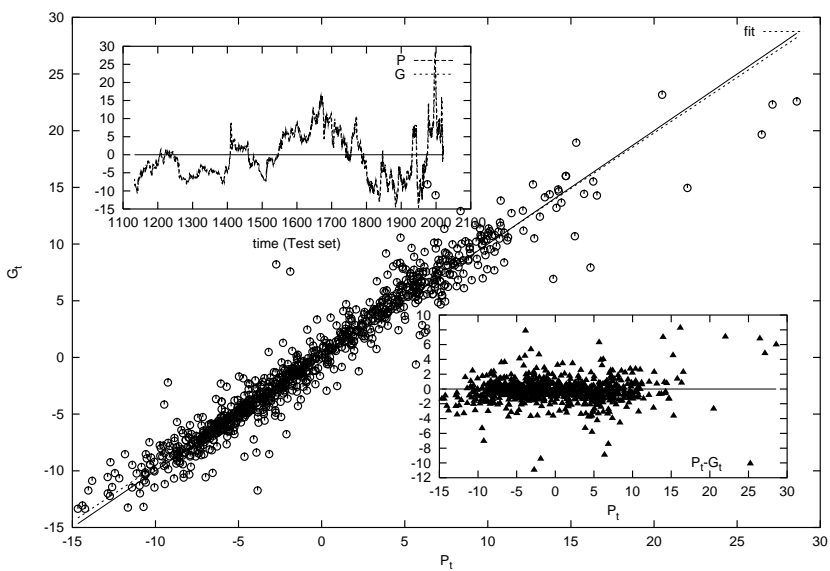,width=0.9\textwidth}
  \caption[XXXXx]{
     Forecast of the time series AAPL. Price is expressed in US\$.
     A perfect forecast will be represented by dots on the $y=x$ line
     (shown as the continuous line).
     The dashed line is a linear fit of the points $\{P_t,G_t\}$.
     A raw measure of the error in forecasting is given by the angular
     coefficient of the fitting line. Values close to one indicate $G_t \simeq P_t$.}
    \label{finset apl}
 \end{center}
\end{figure}

The plot in figure \ref{finset apl} compares the forecasted $G_t$ 
and the real $P_t$ values for the time series of Apple Corp. on the test set $T$. 
It also shows a linear fit for the points $\{P_t,G_t\}$.
A raw measure of performance on the test set $T$ can be obtained 
by the slope of the fitting line (let's call it $\theta$).
It will be a value close to one if the fit corresponds to the $y=x$ line, 
i.e., if $P_t=G_t$.
We obtained the following $\theta$'s for the time series in table \ref{tindices} and
\ref{tstocks}:
$\theta_{\hbox{\piccolo S\&P500}}   = 0.906 $,
$\theta_{\hbox{\piccolo DJI}}       = 0.874 $,
$\theta_{\hbox{\piccolo Nasdaq100}} = 0.860 $.
$\theta_{\hbox{\piccolo AAPL}}      = 0.976 $,
$\theta_{\hbox{\piccolo T}}         = 0.921 $,
$\theta_{\hbox{\piccolo AMD}}       = 0.914 $,
$\theta_{\hbox{\piccolo STM}}       = 0.885 $,
$\theta_{\hbox{\piccolo HON}}       = 0.885 $,
$\theta_{\hbox{\piccolo INTC}}      = 0.874 $,
$\theta_{\hbox{\piccolo CSCO}}      = 0.860 $,
$\theta_{\hbox{\piccolo WCOM}}      = 0.847 $,
$\theta_{\hbox{\piccolo IBM}}       = 0.842 $,
$\theta_{\hbox{\piccolo ORCL}}      = 0.824 $,
$\theta_{\hbox{\piccolo MSFT}}      = 0.803 $,
$\theta_{\hbox{\piccolo SUNW}}      = 0.774 $,
$\theta_{\hbox{\piccolo DELL}}      = 0.692 $,
$\theta_{\hbox{\piccolo QCOM}}      = 0.488 $.

\vskip0.1cm\noindent
The final estimation of the performance in forecasting is made by means of 
the \emph{one-step sign prediction rate} $\S$ defined on $T$ as follows
\begin{equation}
	\S = \frac{1}{|T|} \sum_{t\in T} HS( \Delta P_t \cdot \Delta G_t ) 
                                         + 1 - HS(|\Delta P_t|+|\Delta G_t|)
\label{eqS}
\end{equation}
where $\Delta P_t = P_t-P_{t-1}$ the price change at time step $t\in T$
and $\Delta G_t = G_t-P_{t-1}$ is the \emph{guessed} price change at the same time step.
Note that we assume to know the value of $P_{t-1}$ to evaluate $\Delta G_t$.
$HS$ is a modified 
\footnote{
The usual $HS$ function gives 1 in zero, i.e., $HS(0)=1$
} Heaviside function $HS(x)=1$ for $x>0$ and 0 otherwise.
The argument of the summation in eq(\ref{eqS}) gives one only if $\Delta P_t$ and 
$\Delta G_t$ are
non-zero and with same sign, or if $\Delta P_t$ and $\Delta G_t$ are both zero. 
In other words $\S$ is the probability of a correct guess on the sign of the price 
increment estimated on $T$.

In the lower-right inset of figure \ref{finset apl}
it is shown $P_t-G_t$ as function of $P_t$.
One can see that the difference between the real and the forecasted values
clusters for small $P_t$.
Another way see it is to look at the histogram of $\S$ as function of $\Delta P_t$.
In other words the rate of correct guesses on the sign of the price increment
relative to the magnitude of the fluctuation of the real price.
\begin{figure}[htbp] 
\begin{center}
  \leavevmode
  \epsfig{file=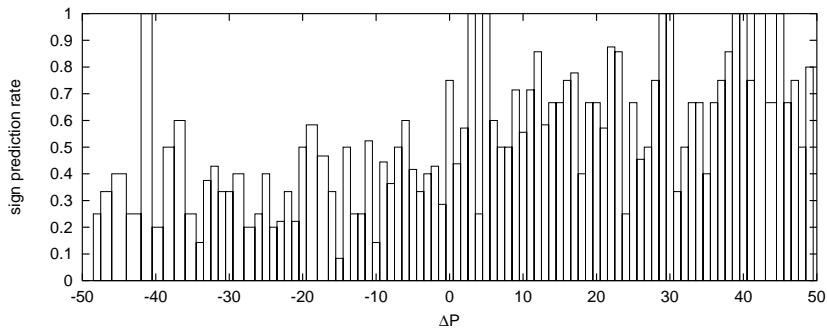,width=0.9\textwidth}
  \caption[jhfg]{
    Normalized $\S$ as function of $\Delta P$ (arbitrary units). 
    The the sign prediction rate
    seems independent from the magnitude of the price change $|\Delta P|$.}
    \label{fhi}
  \end{center}
\end{figure} 
To obtain an unbiased histogram we have to normalize it dividing each bin by the
corresponding value of the $\Delta P$'s histogram (the limit of $\Delta P$ follows
a power law so that large fluctuations are much less probable).
The resulting distribution is plotted in figure \ref{fhi}. It is now clearly visible
that the net does not favor large increments over small ones or vice versa.
In fact the probability to make a correct guess on the sign of the
increment seems independent from the magnitude of the increment itself.
This does not means that the net forecasts ``rare events'' (i.e., a profit opportunity)
as easily as normal fluctuation, because the statistics here
calculated are not significant with respect to extreme events.

To interpret the results that we are going to show we have to concentrate our attention
on the way we select a good net to be used to make forecast. For each time series we
have performed a search to determine the topology of a good net as specified in the
last section. Once we get a pool of candidates the question is ``how many of them
give a sign prediction rate above fifty percent?''

This question is answered in table \ref{toknets}. 
There, \emph{tot} indicates the number of nets such that $\epsilon\le 0.015$, that is, 
we judged as good nets, while \emph{ok} is the number of them that gave 
$\S \ge 50$. This ratio can be seen as an estimation of the confidence that
the net will perform a ``sufficient'' forecast of price change, where sufficient
means above fifty percent.

\begin{table}[tbp] 
\caption[table]{
      Here \emph{tot} indicates the number of nets such that $\epsilon\le 0.015$, 
      that is, we judged as good nets, while \emph{ok} is the number of them that 
      gave a sign prediction rate $\S$ above 50 percent. 
  }
\begin{center}\leavevmode\small
\begin{tabular}{|l l || l l |}\hline
Series      & $ok/tot$ &  Series  & $ok/tot$ \\
\hline\hline
S\&P500     &  32/54     & DowJones Ind &  189/450  \\
Nasdaq 100  &  45/86     &               &           \\
\hline\hline
SUNW        &  112/112   & DELL          &  69/69    \\
WCOM        &  76/76     & AAPL          &  309/311  \\
INTC        &  46/46     & AMD           &  244/245  \\
STM         &  33/269    & ORCL          &  35/35    \\
MSFT        &  21/21     & IBM           &  9/9      \\
CSCO        &  39/48     & HON           &  22/82    \\
T           &  6/6       & QCOM          &  43/436   \\ \hline
\end{tabular}
\end{center}
  \label{toknets}
\end{table} 

The value of $\S$ together with the specification of the number of units per layer
of the best net is reported in table \ref{tindices} and table \ref{tstocks}
along with the dimension of the learning and validation set.

\begin{table}[tbp] 
\caption[table]{
  For each index the net topology $n_i-n_h-1$ is specified along with 
  $\epsilon$, $\S$, $|L|$ and $|V|$. $|T|= 2024 -( n_i+|L|+|V|+|C|)$ and $|C|=200$.
}
\begin{center}\leavevmode\small
\begin{tabular}{lcccccc}\hline
Symbol        & $n_i$&$n_h$  & $|L|$&$|V|$& $\epsilon$ & $\S$(\%)    \\ \hline
S\&P500       &    8&2       & 500  & 300 & 0.008938   & 52.272727   \\ 
DowJones Ind &   13&2       & 700  & 200 & 0.012074   & 51.488423   \\ 
Nasdaq 100    &    4&25      & 700  & 200 & 0.014182   & 50.982533   \\\hline 
\end{tabular}
\end{center}
  \label{tindices}
\end{table} 

\def\bb{$\bullet${ }}
\def\cc{$\circ${ }}
\begin{table}[tbp] 
\caption[tabbbb]{
    Success ratio for the prediction of the sign change.
    For each asset the net topology is specified along with
    $\epsilon$, $\S$ and the number of points in the learning and validation set. 
    In the second column it is specified the symbols from the respective stock exchange 
    NYSE($\circ$) or Nasdaq($\bullet$).
 }
\begin{center}\leavevmode\small
\begin{tabular}{lccccccc}\hline
Company              & Symbol & $n_i$&$n_h$ & $|L|$&$|V|$& $\epsilon$ & $\S$(\%)    \\ 
\hline
\bb Sun Microsys     & SUNW   &      9&7    & 500  & 300 & 0.014435   & 54.005935   \\ 
\bb Dell Computer    & DELL   &      4&18   & 500  & 300 & 0.004315   & 53.543307   \\ 
\bb Mci Worldcom     & WCOM   &      3&2    & 500  & 300 & 0.004024   & 53.392330   \\ 
\bb Apple Comp Inc   &AAPL    &      5&17   & 700  & 300 & 0.013786   & 53.374233   \\ 
\bb Intel Corp       &INTC    &      6&6    & 500  & 300 & 0.009953   & 53.254438   \\
\cc Adv Micro Device &AMD     &      4&23   & 500  & 300 & 0.012339   & 52.952756   \\
\cc ST Microelectron &STM     &      6&2    & 500  & 300 & 0.003978   & 52.465483   \\
\bb Oracle Corp      &ORCL    &      6&2    & 500  & 300 & 0.006333   & 52.366864   \\
\bb Microsoft Cp     &MSFT    &      10&4   & 500  & 300 & 0.008327   & 52.277228   \\
\cc Intl Bus Machine &IBM     &      10&6   & 500  & 300 & 0.006642   & 52.079208   \\
\bb Cisco Systems    &CSCO    &      4&14   & 500  & 300 & 0.008364   & 51.968504   \\
\cc Honeywell Intl   &HON     &      8&2    & 600  & 200 & 0.008506   & 51.877470   \\
\cc AT\&T            &T       &      3&22   & 500  & 300 & 0.014920   & 51.327434   \\
\bb Qualcomm Inc     &QCOM    &      4&25   & 500  & 300 & 0.009888   & 50.295276   \\ 
\hline
\end{tabular}
\end{center}
    \label{tstocks}
\end{table} 

The sign prediction rates range from 50.29\% to 54\%. While the smallest 
values 50.29 may be questionable, the larger values above 54 seem a clear 
indication that the net is not behaving randomly. 
Instead it has captured some regularities in the nonlinearities of the series.

A quite direct test for randomness can be done computing the probability that 
such forecast rate can be obtained just by flipping a coin to decide the next price 
increment. For this purpose we use a random walk ($pr(\hbox{up})=pr(\hbox{down})=1/2$) 
as forecasting strategy $G_{rw_t}$ and observe how many, over 1000 different random 
walks, give a sign prediction rate $\S_{rw}$ defined in eq(\ref{eqS})  
above the value obtained with our net. 
Note that each random walk perform about 1000 time
steps, the same as $|T|$ for that specified time series 
(see table \ref{tindices} and \ref{tstocks}).
These values are reported in table \ref{nullhyp}.
They indicate that except for QCOM the random walk assumption cannot give the same 
prediction rate as the neural net 
\footnote{
In other words, given a neural net which produce $\S$ as prediction rate over 
a certain time series $P_t$ we may compute the probability at which the 
\emph{null hypothesis of randomness} is rejected.
We use a random walk ($pr(\hbox{up})=pr(\hbox{down})=1/2$) as forecasting
strategy $G_{rw_t}$ and then compute 
$\S_{rw}$ defined in eq(\ref{eqS}) on the time series $P_t$. 
The random variable $\S_{rw}$ have mean 0.5 and standard deviation 
$\sigma_{\S_{rw}}$. 
By definition $\S_{rw}$ is the sample mean of $T$ i.i.d. Bernoullian 
random variables. Thus, assuming that $\S_{rw}$  
converges to a Gaussian ${N}{(1/2,\sigma_{\S_{rw}})}$, we can estimate 
the unknown variance of $\S_{rw}$ as 
$\hat{\sigma}^2_{rw} =  1/N \sum_{i=1}^{N} (\S_{rw_i} - 1/2)^2$.
To have an estimation of $\sigma_{\S_{rw}}$ we ran $N=1000$ random walks each 
giving a value for $\S_{rw}$. 
Once we estimate $\sigma_{rw}$, the null 
hypothesis becomes ``what is the probability $P_{\S_{rw}}[x > \S]$ that 
the neural net is doing a random prediction on $P_t$ with rate $\S$ ?'' 
or the other way around
"what is the probability $P_{\S_{rw}}[x \le \S]$ that the net is not doing 
randomly?''.  Formally 
$
  P_{\S_{rw}}[x\le \S] = 
  \int_{-\infty}^{\S} {N}{({1\over 2},\hat{\sigma}_{rw})}(x)dx 
$
where ${N}{({1\over 2},\hat{\sigma}_{rw})}$ is a Gaussian 
and $\hat{\sigma}_{rw}$ is the estimation of the standard deviation $\sigma_{rw}$ of 
the random variable $\S_{rw}$.
In summary, for every sign prediction rate $\S$ obtained with our neural net on 
a time series $P_t$, we first estimate $\hat{\sigma}_{rw}$ as specified above, then
we compute the probability $P_{\S_{rw}}[x\le \S]$ at which the null
hypothesis of randomness prediction is rejected.
The results tell us that for some bad prediction values (like for QCOM or Nasdaq100)
the randomess hyphothesis cannot be rejected but for the majority of the series
the probability to reject the null hypothesis is something between 0.01 and 0.1.
}.
\begin{table}[tbp] 
\caption[TT]{
     For every sign prediction rate $\S$ reported in table \ref{tindices} and 
     \ref{tstocks} it is here shown the number of random walks (over 1000) 
     that have totalized a sign prediction rate $\S_{rw}$ greater or equal $\S$.
     }
\begin{center}\leavevmode\small
\begin{tabular}{|l l || l l |}\hline
Series      & $\#rw:{\S_{rw}\ge\S}$ &  Series   & $\#rw:{\S_{rw}\ge\S}$ \\
\hline\hline
S\&P500     &  78   & DowJones Ind &  186  \\
Nasdaq 100  &  258  &               &       \\
\hline\hline
SUNW        &  7    & DELL          &  16   \\
WCOM        &  13   & AAPL          &  25   \\
INTC        &  21   & AMD           &  30   \\
STM         &  50   & ORCL          &  69   \\
MSFT        &  76   & IBM           &  103  \\
CSCO        &  98   & HON           &  108  \\
T           &  194  & QCOM          &  431  \\ \hline
\end{tabular}
\end{center}
     \label{nullhyp}
\end{table} 
%

\section{Weekly and intra-day data}

It is interesting to ask if the MLP may exploit regularities in the time series
of price sampled at a lower/higher rate than daily.
Apart from the ``scaling behaviour'' observed empirically in real price series
we are interested in the performance of our procedure (search plus learn) when 
we change the time scale on which we sample the price of the assets or the index at 
a stock market.

To answer this question we performed the same search for the good net on the
IBM and AMD stock price sampled on weekly basis as well as taking intra-day data with
the frequency of one minute.
Both series consisted of 2024 points, the same as the daily price series.

The outcome is that intra-day data are much difficult to be forecasted with our MLPs.
In fact for both the one-minute-delay data series the search did not succeeded 
to find a good net; all the good nets (few) 
have given a sign prediction rate $\S<40\%$.

On the other hand the forecast of weekly data gave a success rate comparable with
that of daily series (e.g., a $4$-$2$-1 net performed $\S = 51.422764$ with 
$\epsilon=0.004947$).

\section{Artificially generated price series}

As last question, and to further test the correctness of our prediction, we
tried to forecast the sign of price changes of an artificially generated
time series. This was generated by the the Cont-Bouchaud herding
model that seems one of the simplest one able to show fat tails
in histogram of returns \cite{Cont:98}.
This model shows the relation between the excess kurtosis observed in the
distribution of returns and the tendency of market participants to imitate
each other (what is called \emph{herd behaviour}).
The model consists of percolating clusters of agents \cite{Stauffer}.
At a given time step a certain number of coalitions (clusters) decide what to do:
they buy with probability $a$, sell with probability $a$ or stay inactive with 
probability $1-2a$. 
The demand of a certain group of traders is proportional to its size and the
total price change is proportional to the difference between supply and demand.

It is clear that such a model generates unpredictable time series, and our
networks should not be able to make any predictions.
Indeed, when our method was 
applied to this series it did not succeeded to find a good net as all the 
tried nets performed bad on the check set $C$, i.e., $\epsilon > 0.015$.

\label{discussion}
\section{Discussion}

We have shown that a suitable neural net able to forecast the sign of the 
price increments with a success rate slightly above 50 percent on a daily 
basis can be found. 
This can be an empirically demonstration that a good net exists but we do not 
have a mechanism to find it with ``high probability''.
In other words we cannot use this method as a profit opportunity because we do 
not know \emph{a priori} which net to use.
Perhaps a better algorithm to search for the good topology 
(model selection and pruning with sensitivity analysis \cite{LNCS:1524:98})
would give some help.
The future work will likely undertake this direction.

As final remark we have found that intra-day data are much more difficult to be 
forecasted with our method than daily or weekly.

\vskip0.5cm
\paragraph{Acknowledgements:}
The author wishes to acknowledge D.~Stauffer and 
G.H.~Zimmermann for useful comments and hints.



\begin{thebibliography}{} 

\bibitem[\protect\citeauthoryear{Weigend and Gershenfeld}{1994}]{vastbib}
A.S.~Weigend and N.A.~Gershenfeld editors., (1994),
\emph{Time Series Prediction: Forecasting the Future and Understanding the Past},
Reading, MA: Addison-Wesley.

\bibitem[\protect\citeauthoryear{Campbell, Lo and MacKinlay}{1997}]{Campbell:97}
J.Y.~Campbell, A.W.~Lo, A.C.~MacKinlay, (1997),
\emph{The Econometrics of Financial Markets},
Princeton Univ. Press.

\bibitem[\protect\citeauthoryear{Moody and Neuneier + Zimmermann}{1998}]{LNCS:1524:98}
J.~Moody, \emph{Forecasting the Economy with Neural Nets: A survey of Challenges 
and Solutions} and R.~Neuneier, H.G.~Zimmermann, \emph{How to Train Neural Networks},
in \emph{Neural Networks: tricks of the trade}, 
edited by Genevieve B. Orr and Klaus-Robert M\"uller, (1998),
Lect. N. Comp. Sci 1524, Springer Heidelberg.

\bibitem[\protect\citeauthoryear{Refenes, Burgess and Bentz}{1997}]{Refenes:97}
A.P.N.~Refenes, A.N.~Burgess and Y.~Bentz, (1997), \emph{Neural Networks in Financial
Engineering: A Study in Methodology}, IEEE Transactions on Neural
Networks {\bf 8(6)}.

\bibitem[\protect\citeauthoryear{White}{1992}]{White:92}
H.~White, (1992), \emph{Artificial Neural Networks approximation and Learning Theory},
Blackwell Publishers, Cambridge, MA.

\bibitem[\protect\citeauthoryear{Cybenko}{1989}]{Cybenko:89}
G.~Cybenko, (1989), \emph{Approximation by superposition of a sigmoidal function},
Mathematics of Control, Signal and Systems, {\bf 2}, 303-314.

\bibitem[\protect\citeauthoryear{Rumelhart, Hinton and Williams}{}]{Rumelhart:86a}
D.E.~Rumelhart, G.E.~Hinton and R.J.~Williams,
\emph{Learning internal representation by Error Propagation}, in \emph{Parallel
Distributed Processing: Exploration in the Microsctructure of Cognition.
Volume I: Foundations}, edited by D.E.~Rumelhart and J.L.McClelland, 318-362,
Cambridge, MA: MIT Press/Bradford Books.

\bibitem[\protect\citeauthoryear{Rumelhart, Hinton and Williams}{1986}]{Rumelhart:86b}
D.E.~Rumelhart, G.E.~Hinton and R.J.~Williams, (1986),
\emph{Learning representation by back-propagation error}, 
Nature, {\bf 323}, pp. 533-536.

\bibitem[\protect\citeauthoryear{Press, Teukolsky, Vetterling and Flannery}{1994}]{Press:94}
W.H.~Press, S.A.~Teukolsky, W.T.~Vetterling and B.P.~Flannery, (1994),
{\it Numerical recipes in C: the art of scientific computing},
Cambridge University Press.

\bibitem[\protect\citeauthoryear{Cont and Bouchaud}{1999}]{Cont:98}
R.~Cont and J.P.~Bouchaud, (1999),
\emph{Herd behaviour and aggregate fluctuations in financial markets},
Macroeconomic Dynamics, in press,
{\tt cond-mat/9712318}. 

\bibitem[\protect\citeauthoryear{Stauffer}{2000}]{Stauffer}
D.~Stauffer, (2000), in proceeding of 
``Economic dynamics from the physics point of view'',
Physics Center Bad Honnef, Germany, March 27-30, 2000, Physica A, in press. 



\end{thebibliography}

\label{lastpage}
\end{document}